\documentclass[final,5p,times,twocolumn]{elsarticle}

\usepackage{amssymb}
\usepackage{amsmath,amssymb,amsfonts,mathrsfs}
\usepackage{algorithmic}
\usepackage[linesnumbered,boxed,ruled,vlined]{algorithm2e}
\usepackage{caption}
\usepackage{ragged2e}

\journal{Advanced Engineering Informatics}

\begin{document}

\begin{frontmatter}



\title{Explainable AutoML (xAutoML) with adaptive modeling for yield enhancement in semiconductor smart manufacturing}


\author[th]{Weihong Zhai}
\ead{zhaiwh@shu.edu.cn}

\author[th,ntu]{Xiupeng Shi\corref{corl}}
\ead{sxp@shu.edu.cn}
\cortext[corl]{Corresponding author}

\author[ntu]{Yiik Diew Wong}
\ead{cydwong@ntu.edu.sg}

\author[th]{Qing Han}
\ead{hanq@shu.edu.cn}

\author[th]{Lisheng Chen}
\ead{shuaytgah@gmail.com}

\affiliation[th]{organization={Shanghai University (SHU)},
            city={Shanghai},
            postcode={201800}, 
            country={China}}

\affiliation[ntu]{
organization={Nanyang Technological University},
            postcode={639798}, 
            country={Singapore}}

\begin{abstract}
Enhancing yield is recognized as a paramount driver to reducing production costs in semiconductor smart manufacturing. However, optimizing and ensuring high yield rates is a highly complex and technical challenge, especially while maintaining reliable yield diagnosis and prognosis, and this shall require understanding all the confounding factors in a complex condition. This study proposes a domain-specific explainable automated machine learning technique (termed xAutoML), which autonomously self-learns the optimal models for yield prediction, with an extent of explainability, and also provides insights on key diagnosis factors. The xAutoML incorporates tailored problem-solving functionalities in an auto-optimization pipeline to address the intricacies of semiconductor yield enhancement. Firstly, to capture the key diagnosis factors, knowledge-informed feature extraction coupled with model-agnostic key feature selection is designed. Secondly, combined algorithm selection and hyperparameter tuning with adaptive loss are developed to generate optimized classifiers for better defect prediction, and adaptively evolve in response to shifting data patterns. Moreover, a suite of explainability tools is provided throughout the AutoML pipeline, enhancing user understanding and fostering trust in the automated processes. The proposed xAutoML exhibits superior performance, with domain-specific refined countermeasures,  adaptive optimization capabilities, and embedded explainability. Findings exhibit that the proposed xAutoML is a compelling solution for semiconductor yield improvement, defect diagnosis, and related applications. 
\end{abstract}




\end{frontmatter}


\section{Introduction}
In the semiconductor industry, yield enhancement is a pivotal concern, with direct consequences for cost efficiency and market competitiveness. The semiconductor industry confronts significant challenges in the post-Moore era, as the doubling pace of integrated circuit complexity becomes increasingly untenable. Enhancing yield has emerged as a pivotal lever to curtail costs and amplify financial returns \cite{2021WSCE}, \cite{lee2023semiconductor}. In advanced logic wafer fabrication facilities (fabs), a mere 1\% increase in yield can translate to a substantial \$150 million in additional estimated net profit \cite{TODD}. To address this, machine learning (ML) has been increasingly employed to augment yield enhancement strategies, such as analyzing critical process steps by feature selection \cite{criticalprocesssteps}, assisting in troubleshooting and process optimization by data mining \cite{chien2007data}, \cite{doke2020datamining}, detecting the potential cause of anomalies by clustering algorithms \cite{furnari2021ensembled}, automatic defect classification \cite{588270}, among others. 

Despite the potential of these ML techniques, their development and deployment typically require extensive expertise, presenting a barrier to rapid integration and responsiveness in semiconductor smart manufacturing (SSM). There is a perennial quest to develop measures to quickly adapt to changes, improve product yields, and optimize resource utilization to enable intelligent, efficient, and responsive operations \cite{lu2020smart}.

By integrating multiple ML functions to rapidly solve different yield problems and automated optimizing configurations, automated machine learning (AutoML) aims to provide reliable solutions for yield enhancement with minimal human intervention and computational resources \cite{he2021automl}. Besides, AutoML will also drive the evolution of manufacturing architectures towards integrated networks of autonomous manufacturing with self-adaption, self-configuration, and self-optimization capabilities. However, there are two main intrinsic challenges for deploying AutoML: (1) prevailing AutoML models, such as Alibaba Cloud PAI AutoML \cite{PAI}, Microsoft Azure ML \cite{barnes2015azure}, are designed as general-purpose systems and lack the nuanced response capabilities required for specialized semiconductor domain issues, and (2) AutoML is an end-to-end black box model and lacks explainability \cite{ixautoml}. 

The semiconductor manufacturing process is generally highly complex and requires precise control over numerous steps and variables, which raises many complicated properties and deployment challenges. Besides, for new-generation human-centric smart manufacturing, a reliable AutoML system also needs to provide explanations that support model outputs and mappings from inputs to outputs and obtain more information to determine the cause of low yield \cite{ZHANG2023102121}. Therefore, this study proposes an explainable AutoML (xAutoML) with adaptive modeling to address the challenges of AutoML applications for yield enhancement in SSM, which combines targeted countermeasures and dominant explainable methods. 

The main contributions of this paper can be summarized as follows:

(1) We propose a smart manufacturing solution to achieve self-adaptation, self-optimization, and automation;

(2) We design a domain-specific AutoML framework to enhance yield rate, bringing more accurate results and efficient performance;

(3) We combine mainstream explainable methods to build a more understandable AutoML pipeline and increase the reliability of solutions.

This paper focuses on exploring the development of a domain-specific explainable AutoML in smart manufacturing. 
Section II reviews the literature research on the main challenge and related works.
Section III elaborates on the design methodology of the xAutoML.
Section IV presents the solution analysis to prove the reliability of xAutoML by data-driven insights.
The final two sections cover discussion and conclusions.

\section{Related works}

\subsection{Yield enhancement in semiconductor smart manufacturing}

Compared with other industries, there are additional complex properties often encountered in the semiconductor manufacturing pipeline, making practical development of yield-enhancing machine learning solutions more challenging \cite{criticalprocesssteps}, \cite{conceptdrift}, including:

\textbf{Random sampling in measurement:}  The probability of obtaining complete measurement data for all process steps is extremely low in semiconductor manufacturing, making it difficult to explore their correlation and reflect sufficient information.

\textbf{Concept drift:} Manufacturing process data is often in the form of data streams. Over time, the underlying probability distribution of the data changes due to subtle changes in the process.

\textbf{Small number of low yield dies:} The failure dies only occupy around 2\% of total dies in flash memory data. Thus, the number of failure samples is probably only a small part of the sample set.

\textbf{High rate of missing values:} Missing data occurs frequently because of faults in the sensor, data storage, and communication, leading to reduced yield, quality, and productivity. 

\subsection{Explainability for AutoML}
As more black-box models are increasingly used to make important predictions for yield enhancement in SSM, the need for explainability is increasing.
However, the literature shows that there is not yet a common understanding of what level of explainability needs to be achieved \cite{arrieta2020explainable}, \cite{AEI-XAI}.
The most agreed-upon thought of explainability is designed to ensure that the system can make overall decisions and reasoning that humans can understand \cite{XRL202474}, \cite{XAI2022381}. It is also the main goal of building a reliable xAutoML system.
Therefore, the following mainstream explainable methods are combined to construct an understandable xAutoML pipeline \cite{ixautoml}, \cite{zoller2023xautoml}. 

\textbf{Importance analysis of hyperparameters and features:}
Assessing which hyperparameters and features are globally important to improve the performance of ML systems \cite{HyperparameterImportance}.

\textbf{Automated ablation study (AAS):} Evaluating which changes were important to achieve the observed performance improvement compared to the original configuration if an AutoML tool was started with a given configuration (e.g., defined by new loss function or features) \cite{autoablation}.

\textbf{Visualization of hyperparameter effects:} Visualizing the effect of changing hyperparameter settings locally and globally \cite{moosbauer2021explaining}.

\textbf{Visualization of the sampling and optimization process:} Visualizing which areas of the configuration space have an AutoML tool sampled, and which performance can we expect \cite{biedenkapp2019cave}. 

Through the explainability of all elements (e.g., features \& hyperparameters) and the whole sampling process, relying on visual analysis, displaying the optimization and performance efficiency, and other comprehensive information, to achieve visual, adaptive, optimized configuration, the whole AutoML pipeline can be easily understood.

\begin{figure*}[tbp]
\centerline{\includegraphics[width=1\textwidth]{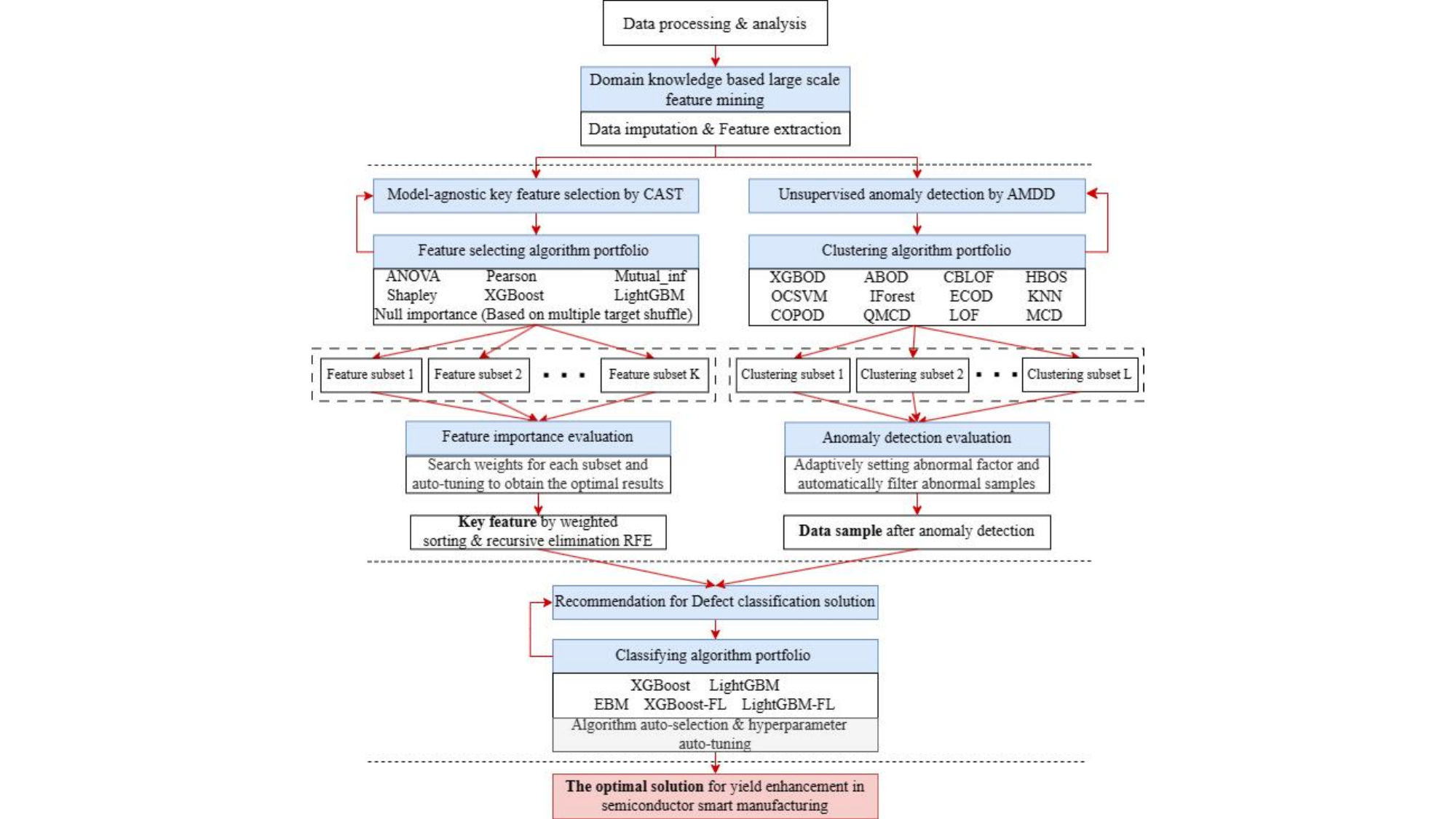}}
\caption{Adaptive xAutoML framework for yield enhancement}
\label{fig1}
\end{figure*}

\subsection{Feature selection methods}
Trillions of production records and billions of daily data increments in petabyte levels are being generated and collected in modern wafer fabs, which contain plenty of signal features. 
However, not all features are equally valuable in a specific semiconductor monitoring system. The original features contain a combination of useless and irrelevant information, which brings significant time costs and inaccurate evaluations \cite{shi2018key}. 
Thus, it is essential to analyze data to identify key components in the process and key factors that affect the response variable.

Feature selection is used to recognize the most important information and reduce computational complexity.
However, due to small changes in data distribution or features, a given selection algorithm may produce wildly different predictions of important features, which makes the results model-specific and inaccurate \cite{shi2020automated}. 
The main reasons are: (1) features are not equally important in different models, which may lead to the Rashomon effect \cite{fisher2019all}; (2) features that are considered highly important in a poor model may not be the same critical features for a well-performing model; and (3) it isn't easy to tune features and model performance iteratively or simultaneously because they are mutual.

\subsection{Class imbalance}  
The occurrence of the failure sample in manufacturing data (i.e., the minority class) is generally much below the norm (i.e., the massive majority class), as described in Section II.A. The issue of extreme differences in prior class probabilities is called class imbalance. Incorrect classification of failure sample into norm entails an enormous misclassification cost \cite{shi2022autocluster}. AutoML in class imbalance is challenging, since algorithms are usually motivated by global optimization, which may be biased towards the majority class, and the minority is likely to be misclassified. 

Several tactics are proposed to reduce the influence of class imbalance, such as data undersampling, oversampling, and cost-sensitive loss functions \cite{diez2015diversity}, \cite{shi2020automated}. However, undersampling can easily lead to loss of useful information, overfitting, and introducing bias. Oversampling can introduce noise or artificially blur the features of the dataset, while insufficient diversity when generating samples can also lead to poor test performance.
To preserve the authenticity of the original data, the targeted loss function enables us to construct adaptive modeling, which typically outperforms the alternatives to sampling heuristic training or minority example mining.

\section{Methodology}

\subsection{xAutoML framework} 
The xAutoML is developed to build an understandable AutoML, which integrates four ML functions commonly used to improve yield into an automatically optimizable pipeline, as shown in Fig.1.

Aims to mine key features and eliminate model dependency, large-scale feature mining based on domain knowledge, and model-agnostic selecting feature technology are proposed, as described in Sections III.B and III.C, respectively. Multiple unsupervised clustering portfolios to detect and evaluate the properties of anomalies, and auto-selection of algorithms for defect classification, are elaborated in Sections III.C and III.D.

Furthermore, we design and deploy an excellent optimization algorithm (termed BOHB) into the whole xAutoML framework by borrowing ideas from a new versatile hyperparametric optimization method \cite{falkner2018bohb}, which uses a probabilistic model Tree-structured Parzen Estimator (TPE) \cite{bergstra2013hyperopt} to guide the search and allocate resources by Bandit-based Hyperparameter Optimization (Hyperband) \cite{li2017hyperband}, to achieve efficient optimization.

\begin{table*}[th]
\caption{\RaggedRight{Variables, imputation, functions, and operations for the large-scale feature mining based on domain knowledge }}
\resizebox{\textwidth}{70mm}{
\begin{tabular}{lll}
\hline
                                                                                    & Code        & Description                                                                                                                                                                                                                                       \\ \hline
\textbf{VARIABLES}                                                                  &             &                                                                                                                                                                                                                                                   \\
Feature                                                                             & x; y        & Measure signal of sub processes($x_i$) and sample ($y_i$)                                                                                                                                                                                                                   \\
Process definition                                                                     & n; c        & Numbers of sub-processes ($n_i$) and cycles of unit process ($c_i$) by data analysis                                                                                                                                                                                       \\
\textbf{DATA IMPUTATION}                                                             &                                                                              &                                                                                                                                                                                                                                                   \\
K-nearest neighbor                                                                   & knn                                                                          &                                                                                                                                                                                                                                                   \\
Mean value                                                                           & mean                                                                         & Converts incomplete data into complete data by replacing the missing values with a mean;                                                                                                                                                          \\
Median value                                                                         & median                                                                       & K-nearest neighbor; median; most frequent value suitable for the data                                                                                                                                                                             \\
Most frequent                                                                        & most\_fre                                                                    & \multicolumn{1}{c}{}                                                                                                                                                                                                                              \\
\begin{tabular}[c]{@{}l@{}}Generative adversarial\\ imputation network\end{tabular} & gain        & \begin{tabular}[c]{@{}l@{}}Observe some components of the real data vector, interpolate the missing components based on what was \\ observed, and output the complete vector by the generator (G) and discriminator (D)\end{tabular} \\
\textbf{FUNCTIONS}                                                                  &             &                                                                                                                                                                                                                                                   \\
Sliding window                                                                      & win5        & Computes the mean, std, max, and min of data with a window size of 5 and a minimum period set to 1                                                                                                                                                 \\
Percentage change                                                                   & pct         & Percentage change of a variable, calculated by $\frac{x_{i+1}-x_i}{x_i}*100$                                                                                                                                                                           \\
Difference                                                                          & diff        & Measures difference of variables, calculated by $x_{i+1}-x_i,y_{i+1}-y_i, c_{i+1}-c_i$                                                                                                                                                        \\
Absolute                                                                            & abs         & Absolute value of columns                                                                                                                                                                                                                         \\
\begin{tabular}[c]{@{}l@{}}Cumulative distribution \\ function\end{tabular}         & cdf         & Describes the cumulative probability of a random variable, calculated by $F_{x}(x) = P(X\textless x)$                                                                                                                                         \\
\begin{tabular}[c]{@{}l@{}}Pearson correlation \\ coefficient matrix\end{tabular}   & corp        & \begin{tabular}[c]{@{}l@{}}Computes pairwise correlation of feature columns or sub-processes in different unit process, \\ calculated by p1 = corr($x_i$, $x_{i+1}$), p2 = corr($n_{c_i}$, $n_{c_j}$)\end{tabular}                                      \\
\begin{tabular}[c]{@{}l@{}}Principal components \\ analysis\end{tabular}            & pca         & Maps multidimensional features to small dimensions by setting thresholds                                                                                                                                                                          \\
\begin{tabular}[c]{@{}l@{}}Independent components \\ analysis\end{tabular}          & ica         & Builds a linear mixing matrix $s_1$,...,$s_n$, by s = $\sum_{i=1}^{n}a_i*x_i$                                                                                                                                      \\
Data filtering                                                                      & filter      & Data smoothing processing                                                                                                                                                                                                                         \\
\begin{tabular}[c]{@{}l@{}}Quartile coefficient \\ of dispersion\end{tabular}       & qcd         & Computes the interquartile distance for a given feature column, e.g., $\frac{0.75*x-0.25*x}{0.75*x+0.25*x}$                                                                                                                                            \\
Log return                                                                          & logr        & Measures relative change on a logarithmic scale, calculated by $\log\frac{x_{i+1}}{x_i}$                                                                                                                                                                              \\
Fourier transform                                                                   & fft         & \begin{tabular}[c]{@{}l@{}}Represents as a linear combination of sine functions in a periodic frequency-domain signal, \\ calculated by $\mathcal{F}|f(t)|= \int_{\infty}^{\infty}f(t)e^{-2\pi\omega it}dt$\end{tabular}                                                                               \\
Fft with data filtering                                                             & fft\_filter & Fourier transform with filter to reduce sine function noise                                                                                                                                                                                       \\
Laplace transform                                                                   & llt         & Transforms a parameter real number t (t $\ge$ 0) into a complex number s, calculated by $\mathcal{L}|f(t)|= \int_{0}^{\infty}f(t)e^{-st}dt$                                                                                                                                                      \\
\textbf{OPERATIONS}                                                                 &             &                                                                                                                                                                                                                                                   \\
Basic center and dispersion                                                         & mean; std   & Mean and standard deviation value                                                                                                                                                                                                             \\
Extreme values                                                                      & min; max    & Minimum, maximum value                                                                                                                                                                                                                       \\
Percentile values                                                                   & q1; q2; q3  & The , 25th, 50th, 75th percentiles to represent data profile and distribution pattern                                                                                                                                                             \\
Mean absolute deviation                                                             & mad         & Measure variability or dispersion                                                                                                                                                                                                                 \\
Profile shape                                                                       & skew; Kurt  & Unbiased kurtosis (kurt) and skew (skew) of feature using Fisher's definition, normalized by n-1                                                                                                                                             \\ \hline
\end{tabular}}
\end{table*}

\subsection{Knowledge-informed massive feature extraction}
The semiconductor manufacturing process mainly includes four major links: silicon wafer manufacturing, integrated circuit design, front-end process, and back-end process. Among them, many “unit processes” with multiple sub-processes require repetitive execution and meet extremely demanding physical property requirements, such as oxidation, photolithography, cleaning, etching, and planarization \cite{criticalprocesssteps}. 

Thus, many key process knowledge and data patterns are hidden in the original data \cite{chien2007data}. 
In the contemporary semiconductor production environment, a large amount of statistical information is often required to monitor the status of the process and determine that the production is in a controlled state to reduce product quality variation \cite{FEextra2022integrated}.
Besides, domain properties such as missing values and random sampling result in much useless and redundant information, and the attribute of concept drift leads to poor transferability. 

Therefore, a large-scale feature mining based on domain knowledge is developed to extract high-quality features (e.g., informative, highly relational, explainable, and non-redundant), which are the foundation for modeling and problem-solving, as well as mining valuable process knowledge and statistical information. The variables, data imputations, functions, and operations design for feature extraction are listed in Table 1. The procedure is elaborated as follows. 

Step 1. A series of variables involving process definition are derived from raw semiconductor manufacturing data analysis, including numbers of sub-processes and cycles of unit process. 

Step 2. Five methods are used to impute missing data and recover the true pattern of the data, e.g., knn, mean, median, most frequent value of data, and gain \cite{yoon2018gain}.

Step 3. Some functions are constructed to extract deep-meaning information and build the contact of multiple variables. Underlying data patterns are explored from numerical transformation in the frequency domain and logarithmic domain, e.g., llt, logr, fft. A series of statistical information involving the production process is used, such as relative change, difference, similarity match, correlation coefficient, and small sliding windows.

Step 4. Some operations are defined to summarize the pivotal information and profile the feature columns, including nodes of data series, statistical descriptions, threshold-based filtering, aggregated or accumulated values, etc.

\begin{algorithm}[th]
\caption{CAST}
\begin{algorithmic}
\STATE \hspace{-16pt}\textbf{}1. Model-agnostic important feature selection:
    \STATE \hspace{-10pt} a. Selection and preprocess of algorithms and features;
        \STATE \hspace{-4pt} 1. Calculate the similarity of different feature subsets;
        \STATE \hspace{-4pt} 2. Select $K$ algorithms with low overlap rates;
        \STATE \hspace{-4pt} 3. Get the top features $f_a$ ranked by $Rv_f$ and store them into the subset $M_a$;
    \STATE \hspace{-10pt} b. Automatic feature selection;
    \FOR{$\text{iteration} = 1,\ldots,N$} 
            \STATE \hspace{-4pt} 1. Set and select weight $W_a$ of each algorithm and feature quantity $Fs$;
            \FOR{$a = 1,\ldots,K$}
                \STATE \hspace{2pt} a. Get the weighted importance score of feature in $M_a$, $Is_a = W_a \cdot Rv_f$;
            \ENDFOR
            \STATE \hspace{-4pt} 2. Integrate all $f_a$ and $Is_a$ into a set;
            \STATE \hspace{-4pt} 3. Get the total weighted score $T_{ws} = \sum_{a=1}^{n=K} (Is_a)$;
            \STATE \hspace{-4pt} 4. Select the top $Fs$ features in $M$ to train;
        \ENDFOR

    \STATE \hspace{-10pt} c. Compare results and find the optimal features configuration ($Fs$, $W_a$, $T_{ws}$);
    \STATE \hspace{-10pt} d. Get model-agnostic features.

\STATE \hspace{-16pt}\textbf{}2. Key feature by RFE:
    \STATE \hspace{-10pt} a. Get the selected $Fs$ features \{Samples with the model-agnostic features\};
    \FOR{$n = Fs, \ldots, 1$}
            \STATE \hspace{-4pt} 1. Permute $n$ times;
            \FOR{$k = 1,\ldots,n$}
            \STATE \hspace{2pt} a. Remove feature $f_k^n$, obtain subset $S_k^{(n-1)}$;
            \STATE \hspace{2pt} b. Re-training model with $S_k^{(n-1)}$;
            \STATE \hspace{2pt} c. Obtain performance $A_k$;
            \ENDFOR        
            \STATE \hspace{-4pt} 2. For the max $\{A_k\}$, eliminate the feature $f_k^{(n)}$;
        \STATE \hspace{-4pt} 3. Keep $n-1$ important features, obtain subset $S^{(n-1)}$ and performance $A^{(n)}$;
    \ENDFOR
    \STATE \hspace{-10pt} b. Select the model-agnostic key features subset $S'(n)$ with the max $\{A^{(n)}\}$.
\end{algorithmic}
\end{algorithm}  

\subsection{Combined algorithm selecting technology}
The given selection algorithm may produce significantly different prediction results for important features, especially in large-scale semiconductor manufacturing data with extracted features, as described in Section II.C.
Thus, a Combined Algorithm Selecting Technology (CAST) is built to eliminate model dependency and select key features. This procedure filters a set of model-specific features and then uses Recursive Feature Elimination (RFE) to discover the optimal subset from the filtered features, as illustrated in Algorithm 1. 
CAST reflects the average intrinsic value of features, comprehensively addressing the problem itself, rather than being conditioned on a pre-specified model.

\subsubsection{Model-agnostic feature selection}
To eliminate model dependencies, $K$ algorithms whose selected feature subsets with a low overlap rate are integrated into an automated pipeline. The feature's relative importance can be assessed by different metrics, such as split weight, mutual information, average gain, etc.
Firstly, some informative and useful features of each algorithm are preliminarily shortlisted with corresponding ranking value ($Rv_f$, the more important the feature, the higher $Rv_f$) and stored in the corresponding subset ($M_a$).

Then, setting the number of iterations, every time CAST automatically searches weight value ($W_a$) for each $M_a$, it gets the weighted importance score of the feature.
Integrate all features of $M_a$ into a set, and get the total weighted score ($T_{ws}$), represented as:

\begin{equation}
    T_{ws} = \sum_{a=1}^{K} (W_a*Rv_f), 
\end{equation}

All features are sorted from large to small according to $T_{ws}$, and then CAST automatically selects the top $Fs$ features for training. By comparing the training results from different $W_a$ and $Fs$, CAST will be optimized to obtain the desired parameter configuration.
The final optimal solution includes model-agnostic features ($f_{Fs}$), quantity $Fs$, the score of weighted importance $T_{ws}$, and the weight of each selection algorithm $W_a$.

If the weight ratio between algorithms does not converge, it indicates that the model has not yet found the optimal solution. In addition, each algorithm has a corresponding weight value. If the weight is small, it proves that the features selected by the algorithm have a model dependency and low contribution to the model-agnostic features of the given data.

\subsubsection{Key feature by recursive elimination}
To further eliminate redundant features, CAST evaluates the performance impact of an individual feature based on the RFE.
The RFE algorithm works by training the model with the model-agnostic features.
At each elimination step, the least important features are eliminated based on the $T_{ws}$, and the model uses the cross-validation accuracy as a metric to retrain the remaining features. The process starts by using all available $f_{Fs}$ and iterates $Fs$ times until all $f_{Fs}$ are eliminated.

A feature is important if it has a considerable performance drop when eliminated. Suppose the performance changes little or even improves after deleting a feature. In that case, the feature is less important because irrelevant features have a minimal impact, while redundant features contribute less due to their high correlation with other more important features. Moreover, the final selected feature can further verify whether our feature extraction method is effective. 

\subsection{Adaptive modeling}
In the semiconductor manufacturing process, the occurrence of failure samples has two reasons. 
One is illogical data caused by sensor sampling, i.e., anomalies; another is defects, which occur during the production process of poor quality and product damage \cite{zhai2023AutoML}.  
Traditional automatic defect classification (ADC) systems identify and classify failure samples relying on manual experience and ignoring errors from anomalies, resulting in inaccurate determination of defect causes and significant production capacity losses \cite{waferdefect}.
Thus, adaptive modeling for defect diagnosis (AMDD) is designed to contain anomaly detection and defect classification and integrate targeted approaches for domain problems, efficiently solving different scientific problems brought by anomalies or defects.

\subsubsection{Automatic clustering for unsupervised anomaly detection}

Anomalies can increase the impact of concept drift, which causes reduced prediction accuracy and poor generalization ability.
Thus, AMDD integrates a set of ($L$) unsupervised clustering algorithms to detect abnormal patterns and avoid significant differences in the concept of clusters \cite{shi2022autocluster}, as shown in Fig.1.

Preprocessing the sample label value $Y$ to ``-1" or ``1", there are $L$ prediction results from clustering algorithms for each sample. We use $N_i$ and $P_i$ to respectively represent the algorithm predicted as ``-1" and ``1". Abnormal factor $A_f$ is used to compare the degree of deviation between the predicted results and the original data, calculated by:

\begin{equation}
    A_f = Y*(\frac{\sum_{i=1}^{L}P_i}{\sum_{i=1}^{L}N_i})
\end{equation}

Thus, a $A_f$ value that is too low in the positive or negative range indicates an abnormality in the normal or failure sample. AMDD automatically detects and filters anomalies in the data by adaptively setting the threshold of $A_f$. By identifying data instances that deviate significantly from the majority of data objects, early anomaly detection is critical to decrease the impact of anomalies that result in yield loss and increase defect classification efficiency.

\subsubsection{Algorithm selection for defect classification based on class imbalance}

Existing defect classification is performed manually using the naked eye in most work sites, which is unreliable and time-consuming. Current AutoML has poor application in building ADC systems because of a list of troublesome properties, such as class imbalance and algorithm selection. To avoid the bias of class imbalance, our model imports the focal loss function, which introduces a modulating factor that downweights the contribution of well-classified examples and focuses more on hard or misclassified examples \cite{lin2017focal}. By assigning higher weights to challenging samples, focal loss helps the model to pay more attention to the minority class and improve overall performance, especially in scenarios where the majority class dominates the dataset.

Focal loss is defined mathematically as:
\begin{equation}
    FL(p_t) = - \alpha_t * (1 - p_t) ^ \gamma * log(p_t).
\end{equation}

Among them, $p_t$ is the probability of the true class. $\alpha_t$ balances the importance of positive and negative samples, and its search interval is set according to the data imbalance rate. 
$\gamma$ is used to reduce the weight of easy-to-classify samples and focus on training difficult-to-classify samples (the larger $\gamma$, the smaller the loss of easy samples, and more attention is paid to difficult samples).

To select the optimal algorithm for defect classification based on class imbalance, AMDD then uses the explainable method AAS to build an automatic pipeline. The pipeline integrates advanced algorithms as baseline models, including Explainable Boosting Machines (EBM) \cite{nori2019interpretml}, Extreme Gradient Boosting (XGBoost) \cite{chen2016xgboost}, LightGBM \cite{ke2017lightgbm}.
Then AMDD performs an ablation operation to change the loss function of XGBoost and LightGBM from cross-entropy loss to focal loss (termed XGBoost-FL \& LightGBM-FL). 
Finally, all hyperparameters of algorithms and focal loss are automatic search and optimization, F1-score is set as the metric to evaluate the results of ablation experiments.
According to the assumption, algorithms with focal loss should have more prominent performance in imbalanced defect classification problems.

\section{Analysis}
\subsection{Data}
As a test scenario, this study used the SECOM dataset from a modern semiconductor manufacturing process, which is available on the UCI machine learning repository and collectible via the monitoring of signals$/$variables from sensors and process measurement points \cite{misc_secom_179}.
The SECOM represents a complex, high-dimensional, and highly imbalanced real-world problem, containing 1,567 samples, each with 590 digital features. 
According to the histogram analysis of the original data, there are 41,951 missing data and 347 feature columns with variance$\textless{}$1, accounting for 58.6\%, of which there are 116 useless columns with the same value.
The data set has labels, ``1" indicates a failed product, and ``-1" indicates a qualified product. There are 104 failure samples and 1,463 normal samples. The data imbalance ratio of this dataset reaches a high value of 14:1.
  
The large-scale feature mining methods based on domain knowledge are used to extract useful information and underlying data patterns from original features. Four unit processes with 128 sub-processes are defined through the periodic distribution pattern of missing values.
As a result, more than 60,000 features are extracted comprehensively from the original 590 features in SECOM, approaching two orders of magnitude.
The extracted feature can serve as a reference and basis for engineers to conduct advanced investigations into process control and capability analysis, and also provides underlying important features for model performance improvement, as described in Section IV.B.

\subsection{Model-agnostic key features selection by CAST}
CAST is developed to select model-agnostic key features from complex and high-dimensional extracted features.
In the manufacturing process, these features can help experts determine key factors contributing to yield excursions downstream and assist in making decisions for possibly identifying the specific data patterns. 
For model performance, the benefits of these features include better explainability, simplified modeling, shorter learning time, and enhanced generalization, among others.
The final feature selection scheme proves CAST eliminates model dependency and performs more accurately, effectively, and explainably compared to the given selection algorithm.

\subsubsection{Weighted selection for model-agnostic features}
As shown in Fig.2(a), during the hyperparameter (e.g., $W_a$, $Fs$) optimization process, the model gradually adjusts the search strategy, and the training effect continues to improve. The maximum performance increases with the number of iterations until the optimal solution is found.
Fig.2(b) shows the change in the proportion of the normalized weight of the selection algorithm corresponding to the current optimal performance. 

Before the $51st$ iteration, CAST tried different weight combinations, and the weights of different algorithms were constantly adjusted. As the iterations increase, the weights of $Mutual\_inf$ and $Pearson$ gradually decrease, while the weights of the other algorithms tend to be stable.
The weight after the $87th$ iteration is approximate with the optimal result in the $183rd$ iteration.

\begin{figure}[tbp]
\centerline{\includegraphics[width=0.5\textwidth]{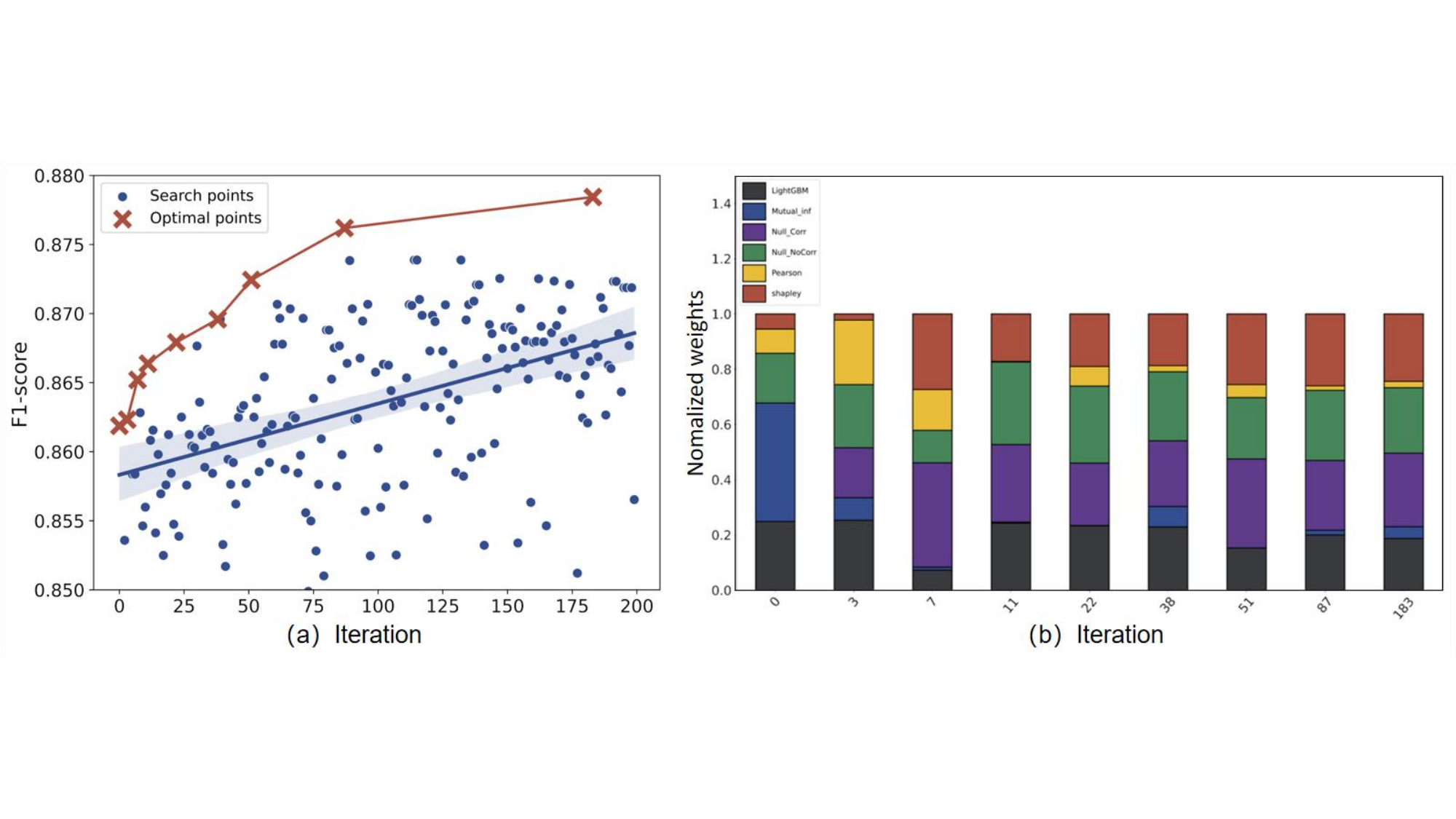}}
\caption{Performance variation trend of CAST(a) and Normalized weight ratio of the current optimal result(b)}
\label{fig2}
\end{figure}

\begin{figure}[tbp]
\centerline{\includegraphics[width=0.5\textwidth]{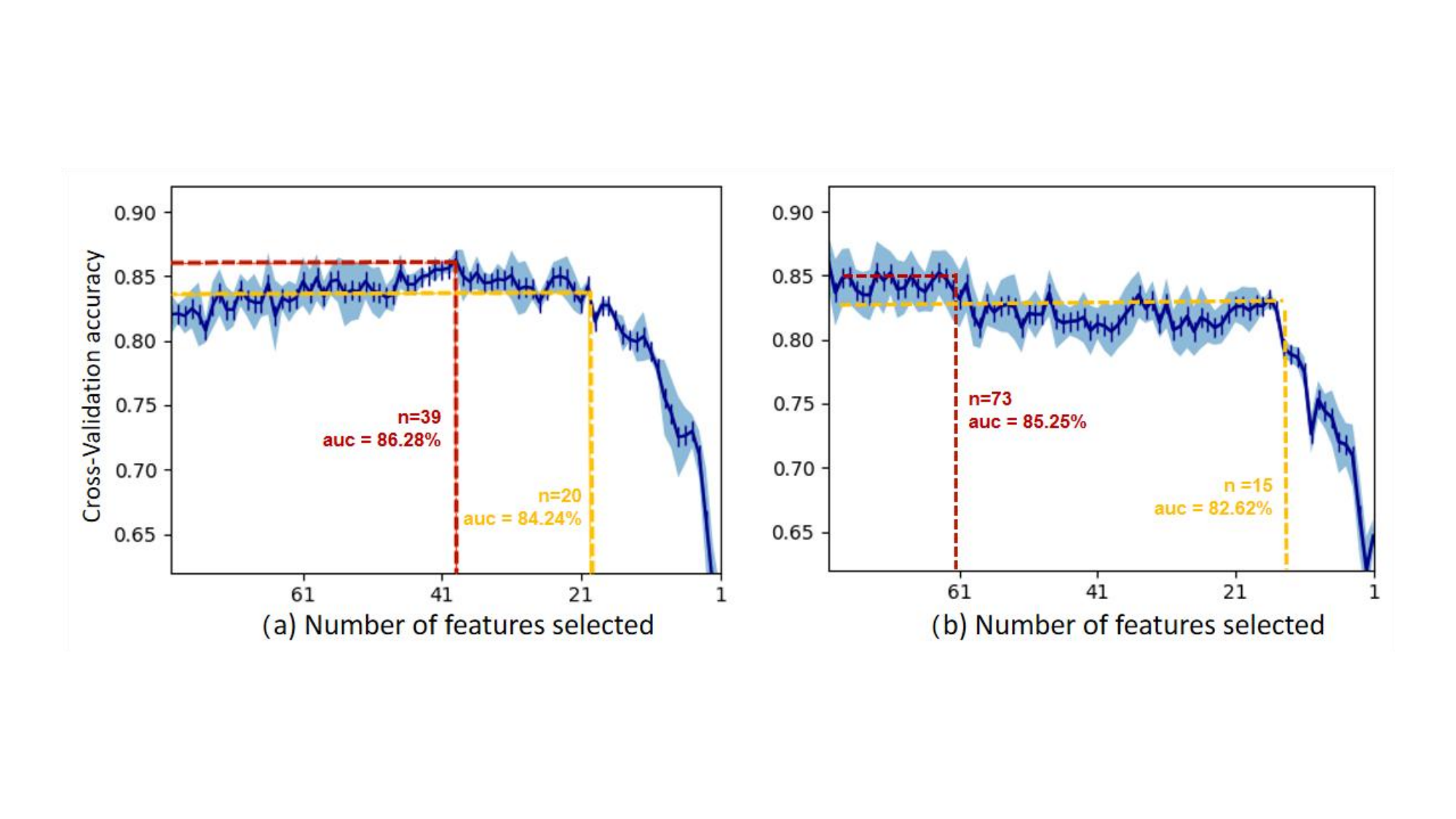}}
\caption{RFE process of CAST(a) and single feature selection method(b)}
\label{fig3}
\end{figure}

These results intuitively explain the importance of different algorithms in the sampling process of weights and features. The weight of different algorithms at the optimal performance is constant, and the goal of hyperparameter optimization is fitting the approximate point of the constant weight.
The convergence of weights and visualization of the sampling process can infer that the model has searched for the optimal solution.

As a result, CAST eliminates model dependency by adjusting the weight between algorithms. The low weight of $Mutual\_inf$ and $Pearson$ indicates that their selected features are model-specific and inaccurate. This is due to their selection based on the statistical distribution pattern of features. 

\subsubsection{Recursive elimination for redundant features}
During the RFE process, key feature combinations are selected from model-agnostic features by model re-training and recursively pruning features with minimal permutation importance in the current set. 

The learning performance of each iteration of CAST is shown in Fig.3(a). Before selecting 39 features, the model's performance gradually increases with the decreased feature, proving that these features are less important.
The performance at the subset with 20 features reaches an accuracy of 84.24\%, but a lower variance. Because the performance has a considerable decrease after the elimination, 39 key features are selected with a cross-validation accuracy of 86.28\%. 

The single feature selection method in Fig.3(b) has two elimination stages. Before the remaining 73 features, the model eliminates redundant features highly correlated with $73rd$, and the performance is maintained at around 85.25\%. After 73 features, some important features are eliminated, causing the model performance to drop to about 82.62\%. Redundant features highly correlated with the $15th$ are eliminated again until there are 15 features left.

By comparison, CAST eliminates more redundant features, obtains more features with equal importance, and brings better performance.
Besides, there are 35 new extracted features of the final feature selection scheme, accounting for 89.7\%, which shows that large-scale feature mining methods are useful and reliable.

\subsection{Detection and evaluation of anomaly}

For anomaly detection, in this experiment, the threshold distributions of $A_f$ in normal samples are set to -5, and failure samples are set to 0.1, which means that the samples of which $A_f\textless{}-5$ and $0\textless{}A_f\textless{}0.1$ are anomalous.

The results show that there are 74 detected anomalous samples all from samples labeled as ``1", which validates the impact of class imbalance. During anomaly detection, the model may be biased towards the majority due to global optimization, and minority classes may be mistakenly ignored as noise or outliers.

In addition, 15 samples are difficult to determine the level of anomalies (11 stems from label ``-1" and 4 stems from label ``1"), which means the number of algorithms that judge samples as abnormal and non-abnormal is the same.
Further identifying the differences analyzing the failure cause of these 15 samples and introducing measures to address class imbalance are the main directions for building efficient anomaly detection methods and optimizing the production line.


\begin{figure}[tbp]
\centerline{\includegraphics[width=0.5\textwidth]{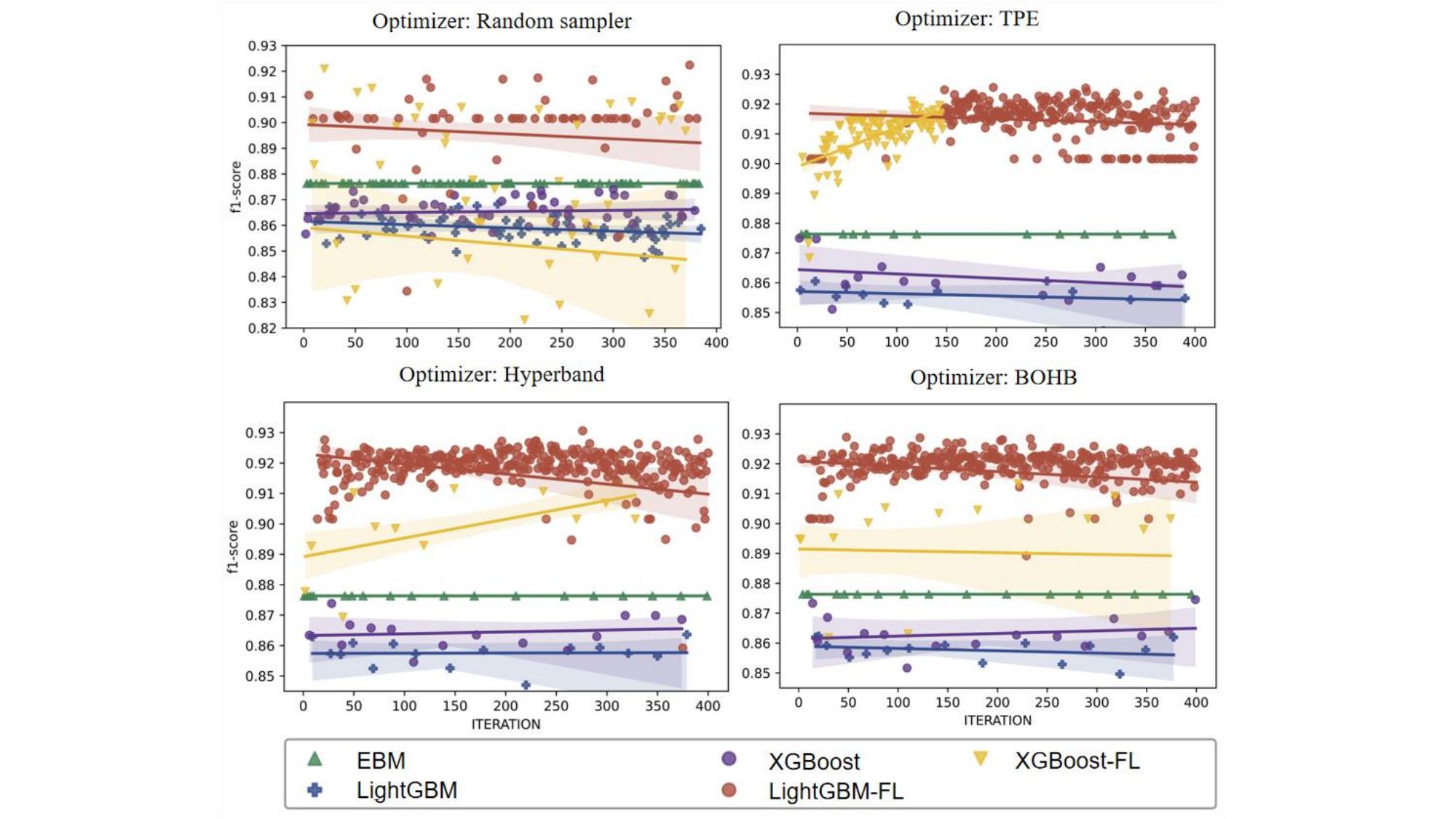}}
\caption{Performance change of classification algorithms by different optimizer}
\label{fig4}
\end{figure}

\subsection{Optimal solution for defect classification}

AMDD has built a fast-deployable, reliable, and efficient ADC system for defect classification by integrating multiple algorithms and focal loss for class imbalance.
Mainstream hyperparameter optimization (HPO) algorithms are used to compare and ensure the superiority of our optimization algorithm.

The sampling process in Fig.4 shows that the algorithm EBM brings better performance gains of 87.75\% than XGBoost and LightGBM. For imbalanced data, algorithms with focal loss have better performance, which improves the performance of XGBoost from 86.5\% to 91\% and increases LightGBM from 85.8\% to 92\%. 

The ablation experiments control loss functions as variables to observe the performance variation trend of different algorithms. 
Due to different search strategies, the sampling process and final performance for each HPO are different. The final optimal algorithm obtained by all HPOs is LightGBM-FL. Our HPO algorithm exhibits the best search performance, achieving an optimization performance of 92.89\% for LightGBM-FL in the shortest iteration time.

These results prove the effectiveness of focal loss and bring a reliable solution for automatic algorithm selection based on class imbalance in SSM. 
The optimal defect classification solution to avoid bias and bring excellent performance is critical for efficiently identifying and fixing product defects, which can reduce production costs and increase yield.

\begin{figure}[tbp]
\centerline{\includegraphics[width=0.5\textwidth]{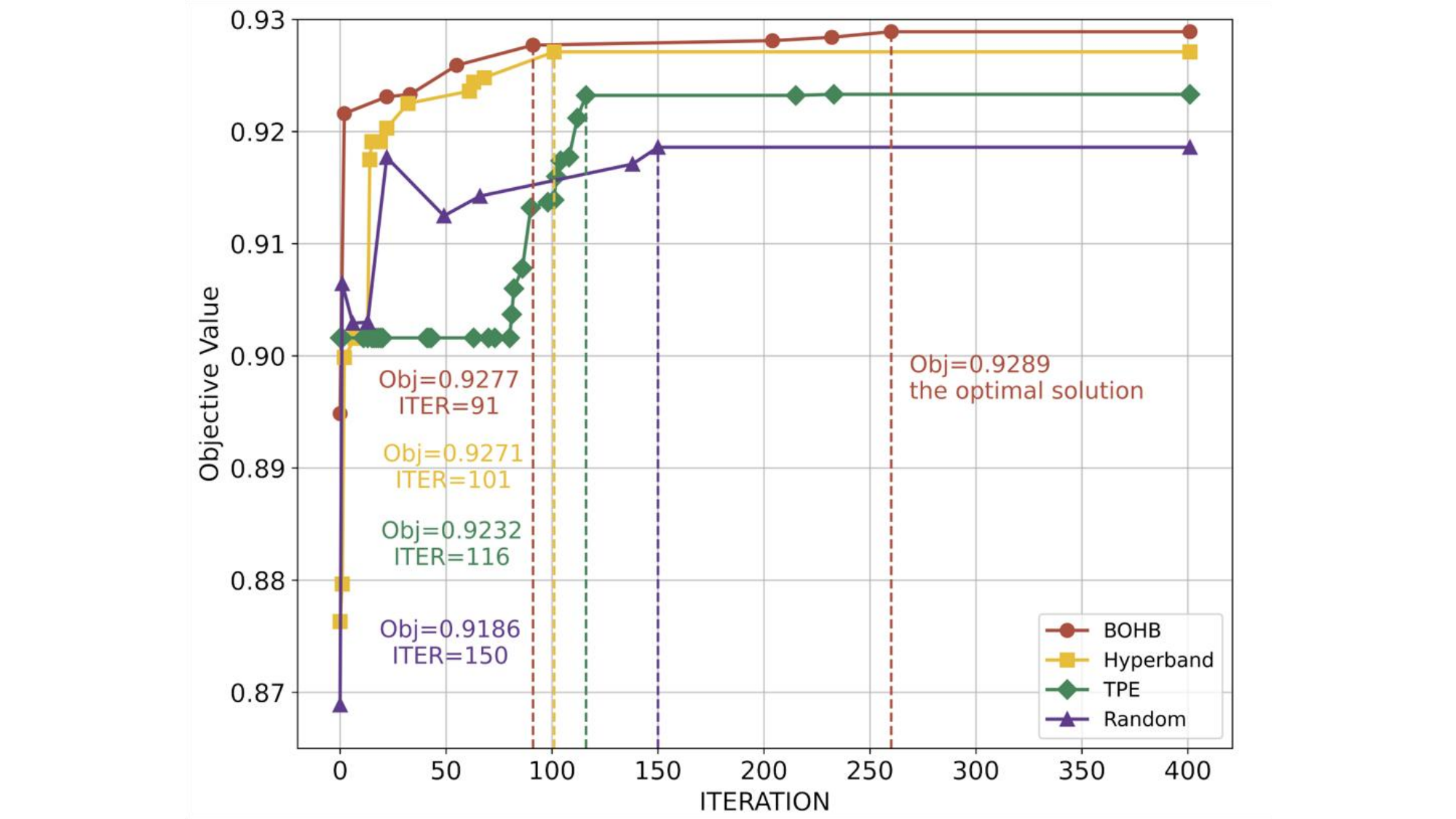}}
\caption{Tuning rules of the maximum performance curves of different optimizers}
\label{fig5}
\end{figure}

\section{Discussions}

\subsection{Optimization performance of HPO algorithms}
To further demonstrate the efficiency of our HPO algorithm, we compare the optimization process of BOHB and other algorithms, as shown in Fig.5.
BOHB shows comprehensive superiority compared with other algorithms.
For the optimization process, the curve of BOHB is globally superior to other algorithms. 
For maximum performance, the ranking of different HPOs is: 92.89\% (BOHB) $>$ 92.71\% (Hyperband) $>$ 92.33\% (TPE) $>$ 91.86\% (Random sampler).
Moreover, BOHB is also the fastest at finding the current optimal solution, and the local performance is far ahead under the same number of iterations.
The number of iterations corresponding to the local optimal performance of all HPOs is ranked as follows: 150 (random sampler) $>$ 116 (TPE) $>$ 101 (Hyperband) $>$ 91 (BOHB).
With a shorter number of iterations and better performance, the results demonstrate the excellence of our optimization algorithm.

\subsection{Limitations and future work}\label{AA1}

\begin{figure}[tbp]
\centerline{\includegraphics[width=0.5\textwidth]{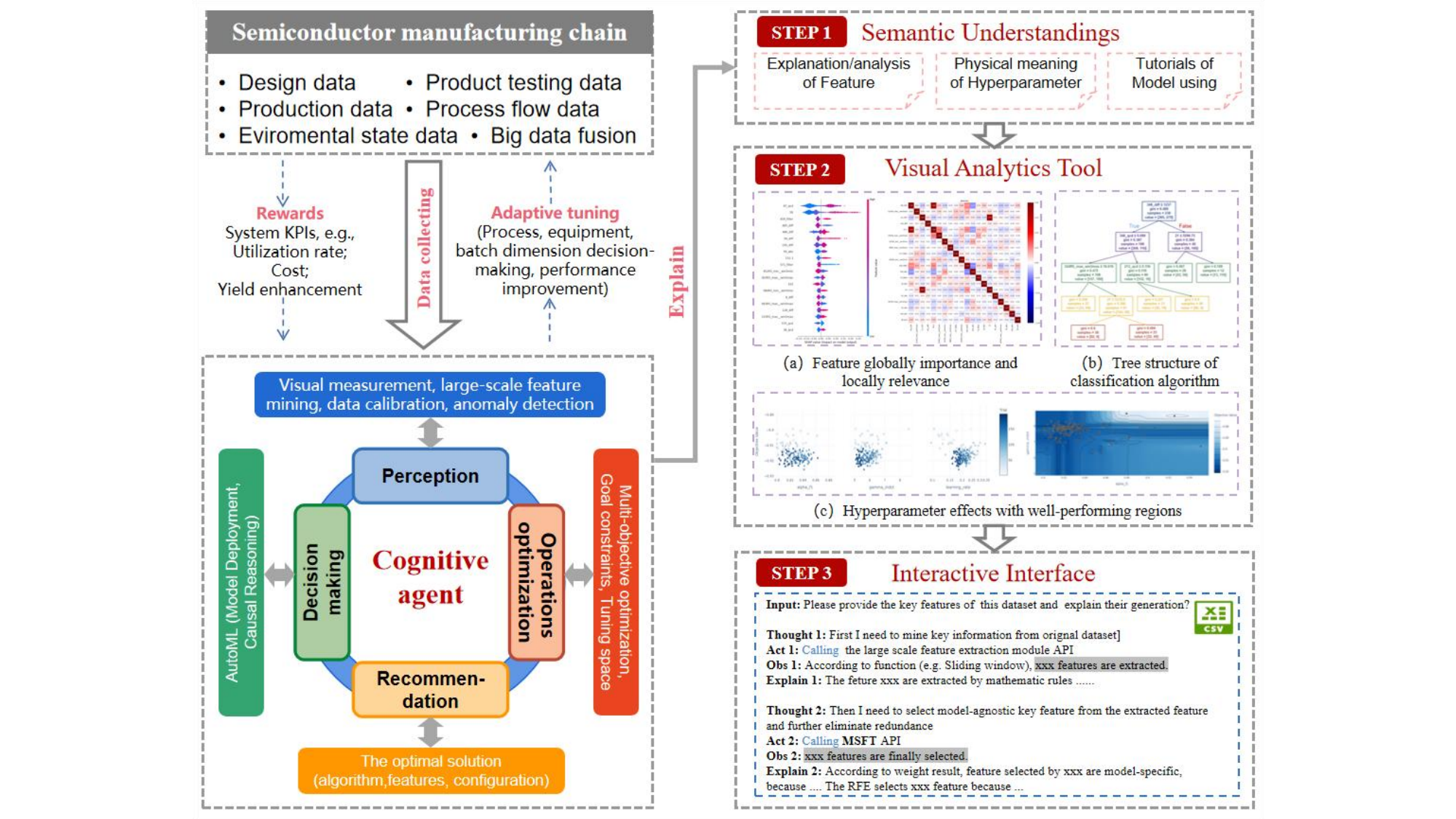}}
\caption{Advanced manufacturing autonomous smart system framework with cognitive decision-making capabilities}
\label{fig6}
\end{figure}
 
Due to the lack of onsite data from semiconductor foundries, the model has not verified the practical feasibility and complete functions.
Besides, the dataset uses signals 1 to 590 to represent encrypted physical meaning for confidentiality concerns.

The xAutoML is currently designed to create post-hoc explanations. Users analyze the results after the optimization is complete. According to the extracted insights, users can deploy targeted measures to increase yield and guide the next optimization run in the correct direction. 

For future work, to further transform the traditional manufacturing pyramid into an advanced manufacturing autonomous intelligent system with cognitive decision-making capabilities, thereby achieving elastic and responsive, explainable SSM operations \cite{lu2020smart}, \cite{moghaddam2018SM}, \cite{shen2006applications}. As a bridge for value transfer between large models and scenarios, AI Agent is autonomous and adaptive, which can learn and improve independently in specific tasks or fields, and each link is explained in detail \cite{AIAgent}. Therefore, we will build cognitive agent manufacturing systems to efficiently solve production scenario problems from the four parts of perception, decision-making, operations optimization, and recommendation, as well as to adaptively optimize manufacturing on the application side and realize SSM \cite{kruger2011agent}.

In addition, the cognitive agent also provides three explainable methods to assist non-domain experts in understanding and hassle-free using the following.

(1) Provide cross-domain connections to create semantic understanding between “things”, for example, physical explanations of features and hyperparameters, as well as providing cases to illustrate the AutoML process, etc.

(2) Set up visual analysis tools to explain key elements of the model. For features, the results in Fig.6(a) reflect the positive and negative impact of their importance on the model output and the degree of correlation (high in red, low in blue). For the algorithm, Fig.6(b) shows the tree structure of its decision, with each node representing a feature and its corresponding threshold. For the impact of hyperparameters, Fig.6(c) visualizes the sampling process and contour plots to establish regions of good and poor performance from the search space, creating custom views for specific areas of interest.

(3) Establish an interactive user interface, perceive and decompose tasks according to input, train the chain-of-thought (CoT) of the model, and synergize reasoning and acting (ReACT) in language models to fit the output desired by the user, thereby calling the pre-trained algorithm module to output the solution and explain the decision-making reasoning process, as well as the specific current data conclusions and results.

\subsection{Application potentials}\label{AA1}
For yield enhancement in semiconductor smart manufacturing, xAutoML has broad application potential and can help the manufacturing industry achieve higher production efficiency, quality control, anomaly detection, and predictive maintenance by defect classification.
Through adaptive modeling and end-to-end optimization, xAutoML automatically selects and recommends the optimal solutions (including key features, defect diagnosis, and algorithm selection), which can help domain experts identify bottlenecks and efficient deployment on the production line and provide optimization suggestions. This allows manufacturers to correct problems, promptly improve production efficiency, and reduce costs.
xAutoML also provides visualization based on internal optimization reasoning and provides a reasonable explanation of results, which balances the performance and explainability of the model well.

\section{Conclusions}
This study designs an integrated approach to build an explainable AutoML for yield enhancement in SSM, which combines targeted measures based on domain knowledge and mainstream explainable methods to bring more accurate results and efficient performance. This research contributes to the manufacturing sphere and associated literature in four fields.

(1) Extraction of massive features from the manufacturing process. We conduct in-depth and multi-perspective measurements to explore key information related to processes and modeling. More than 10,000 features will be extracted comprehensively for a single data imputation method.

(2) Model-agnostic key feature selection by multi-model integration and recursive elimination. 
These high-quality features play a key role in process optimization and model performance. They are chosen using multiple weighted algorithms and the RFE process, and 39 key features are identified. The technology is dependable and effective in identifying important features for yield enhancement and contributing to guiding the orientation of feature engineering, which is the key to improving modeling.

(3) Anomaly detection and rating based on unsupervised clustering. To estimate potential anomalies in manufacturing, provide valuable information for the potential cause, and find the area of the process where the majority of anomalies arise, twelve clustering algorithms were constructed to evaluate the level of anomalies, and four anomaly levels were obtained.

(4) Satisfactory results of defect classification integrated by advanced algorithms and focal loss for class imbalance. Based on the key features in the manufacturing process, product defects were identified, and an overall accuracy of 92.89\% was achieved, demonstrating excellent optimization efficiency. The approach builds a fast-deployable, reliable, and efficient ADC system.


\begin{thebibliography}{58}
\expandafter\ifx\csname natexlab\endcsname\relax\def\natexlab#1{#1}\fi
\providecommand{\url}[1]{\texttt{#1}}
\providecommand{\href}[2]{#2}
\providecommand{\path}[1]{#1}
\providecommand{\DOIprefix}{doi:}
\providecommand{\ArXivprefix}{arXiv:}
\providecommand{\URLprefix}{URL: }
\providecommand{\Pubmedprefix}{pmid:}
\providecommand{\doi}[1]{\href{http://dx.doi.org/#1}{\path{#1}}}
\providecommand{\Pubmed}[1]{\href{pmid:#1}{\path{#1}}}
\providecommand{\bibinfo}[2]{#2}
\ifx\xfnm\relax \def\xfnm[#1]{\unskip,\space#1}\fi
\bibitem[{He et~al.(2021)He, Zhao, and Chu}]{he2021automl}
\bibinfo{author}{X.~He}, \bibinfo{author}{K.~Zhao}, \bibinfo{author}{X.~Chu},
\newblock \bibinfo{title}{Automl: A survey of the state-of-the-art},
\newblock \bibinfo{journal}{Knowledge-Based Systems} \bibinfo{volume}{212} (\bibinfo{year}{2021}) \bibinfo{pages}{106622}.
\bibitem[{Gunning et~al.(2019)Gunning, Stefik, Choi, Miller, Stumpf, and Yang}]{gunning2019xai}
\bibinfo{author}{D.~Gunning}, \bibinfo{author}{M.~Stefik}, \bibinfo{author}{J.~Choi}, \bibinfo{author}{T.~Miller}, \bibinfo{author}{S.~Stumpf}, \bibinfo{author}{G.-Z. Yang},
\newblock \bibinfo{title}{Xai:explainable artificial intelligence},
\newblock \bibinfo{journal}{Science robotics} \bibinfo{volume}{4} (\bibinfo{year}{2019}) \bibinfo{pages}{eaay7120}.
\bibitem[{Ma et~al.(2019)Ma, Wang, Xie, Hong, Mellmann, Sun, Gao, Singh, and Venkatachalam}]{ma2019machine}
\bibinfo{author}{Y.~Ma}, \bibinfo{author}{F.~Wang}, \bibinfo{author}{Q.~Xie}, \bibinfo{author}{L.~Hong}, \bibinfo{author}{J.~Mellmann}, \bibinfo{author}{Y.~Sun}, \bibinfo{author}{S.~W. Gao}, \bibinfo{author}{S.~Singh}, \bibinfo{author}{P.~Venkatachalam},
\newblock \bibinfo{title}{Machine learning based wafer defect detection},
\newblock in: \bibinfo{booktitle}{Design-Process-Technology Co-optimization for Manufacturability XIII}, volume \bibinfo{volume}{10962}, \bibinfo{organization}{SPIE}, \bibinfo{year}{2019}, pp. \bibinfo{pages}{38--45}.
\bibitem[{Lee et~al.(2019)Lee, Yang, Lee, and Kim}]{criticalprocesssteps}
\bibinfo{author}{D.-H. Lee}, \bibinfo{author}{J.-K. Yang}, \bibinfo{author}{C.-H. Lee}, \bibinfo{author}{K.-J. Kim},
\newblock \bibinfo{title}{A data-driven approach to selection of critical process steps in the semiconductor manufacturing process considering missing and imbalanced data},
\newblock \bibinfo{journal}{Journal of Manufacturing Systems} \bibinfo{volume}{52} (\bibinfo{year}{2019}) \bibinfo{pages}{146--156}.
\bibitem[{LeDell and Poirier(2020)}]{ledell2020h2o}
\bibinfo{author}{E.~LeDell}, \bibinfo{author}{S.~Poirier},
\newblock \bibinfo{title}{H2o automl: Scalable automatic machine learning},
\newblock in: \bibinfo{booktitle}{Proceedings of the AutoML Workshop at ICML}, volume \bibinfo{volume}{2020}, \bibinfo{organization}{ICML}, \bibinfo{year}{2020}.
\bibitem[{P{\l}o{\'n}ska and P{\l}o{\'n}ski(2021)}]{plonska2021mljar}
\bibinfo{author}{A.~P{\l}o{\'n}ska}, \bibinfo{author}{P.~P{\l}o{\'n}ski},
\newblock \bibinfo{title}{Mljar: State-of-the-art automated machine learning framework for tabular data},
\newblock \bibinfo{journal}{Version 0.10} \bibinfo{volume}{3} (\bibinfo{year}{2021}).
\bibitem[{Shi et~al.(2019)Shi, Wong, Li, Palanisamy, and Chai}]{shi2019feature}
\bibinfo{author}{X.~Shi}, \bibinfo{author}{Y.~D. Wong}, \bibinfo{author}{M.~Z.-F. Li}, \bibinfo{author}{C.~Palanisamy}, \bibinfo{author}{C.~Chai},
\newblock \bibinfo{title}{A feature learning approach based on xgboost for driving assessment and risk prediction},
\newblock \bibinfo{journal}{Accident Analysis \& Prevention} \bibinfo{volume}{129} (\bibinfo{year}{2019}) \bibinfo{pages}{170--179}.
\bibitem[{D{\'\i}ez-Pastor et~al.(2015)D{\'\i}ez-Pastor, Rodr{\'\i}guez, Garc{\'\i}a-Osorio, and Kuncheva}]{diez2015diversity}
\bibinfo{author}{J.~F. D{\'\i}ez-Pastor}, \bibinfo{author}{J.~J. Rodr{\'\i}guez}, \bibinfo{author}{C.~I. Garc{\'\i}a-Osorio}, \bibinfo{author}{L.~I. Kuncheva},
\newblock \bibinfo{title}{Diversity techniques improve the performance of the best imbalance learning ensembles},
\newblock \bibinfo{journal}{Information Sciences} \bibinfo{volume}{325} (\bibinfo{year}{2015}) \bibinfo{pages}{98--117}.
\bibitem[{L{\'o}pez et~al.(2013)L{\'o}pez, Fern{\'a}ndez, Garc{\'\i}a, Palade, and Herrera}]{lopez2013insight}
\bibinfo{author}{V.~L{\'o}pez}, \bibinfo{author}{A.~Fern{\'a}ndez}, \bibinfo{author}{S.~Garc{\'\i}a}, \bibinfo{author}{V.~Palade}, \bibinfo{author}{F.~Herrera},
\newblock \bibinfo{title}{An insight into classification with imbalanced data: Empirical results and current trends on using data intrinsic characteristics},
\newblock \bibinfo{journal}{Information sciences} \bibinfo{volume}{250} (\bibinfo{year}{2013}) \bibinfo{pages}{113--141}.
\bibitem[{Nori et~al.(2019)Nori, Jenkins, Koch, and Caruana}]{nori2019interpretml}
\bibinfo{author}{H.~Nori}, \bibinfo{author}{S.~Jenkins}, \bibinfo{author}{P.~Koch}, \bibinfo{author}{R.~Caruana},
\newblock \bibinfo{title}{Interpretml: A unified framework for machine learning interpretability},
\newblock \bibinfo{journal}{arXiv preprint arXiv:1909.09223}  (\bibinfo{year}{2019}).
\bibitem[{Rigatti(2017)}]{rigatti2017randomforest}
\bibinfo{author}{S.~J. Rigatti},
\newblock \bibinfo{title}{Random forest},
\newblock \bibinfo{journal}{Journal of Insurance Medicine} \bibinfo{volume}{47} (\bibinfo{year}{2017}) \bibinfo{pages}{31--39}.
\bibitem[{Feurer et~al.(2015)Feurer, Klein, Eggensperger, Springenberg, Blum, and Hutter}]{feurer2015efficient}
\bibinfo{author}{M.~Feurer}, \bibinfo{author}{A.~Klein}, \bibinfo{author}{K.~Eggensperger}, \bibinfo{author}{J.~Springenberg}, \bibinfo{author}{M.~Blum}, \bibinfo{author}{F.~Hutter},
\newblock \bibinfo{title}{Efficient and robust automated machine learning},
\newblock \bibinfo{journal}{Advances in neural information processing systems} \bibinfo{volume}{28} (\bibinfo{year}{2015}).
\bibitem[{Bergstra and Bengio(2012)}]{randomsearch}
\bibinfo{author}{J.~Bergstra}, \bibinfo{author}{Y.~Bengio},
\newblock \bibinfo{title}{Random search for hyper-parameter optimization.},
\newblock \bibinfo{journal}{Journal of machine learning research} \bibinfo{volume}{13} (\bibinfo{year}{2012}).
\bibitem[{Li et~al.(2017)Li, Jamieson, DeSalvo, Rostamizadeh, and Talwalkar}]{li2017hyperband}
\bibinfo{author}{L.~Li}, \bibinfo{author}{K.~Jamieson}, \bibinfo{author}{G.~DeSalvo}, \bibinfo{author}{A.~Rostamizadeh}, \bibinfo{author}{A.~Talwalkar},
\newblock \bibinfo{title}{Hyperband: A novel bandit-based approach to hyperparameter optimization},
\newblock \bibinfo{journal}{The Journal of Machine Learning Research} \bibinfo{volume}{18} (\bibinfo{year}{2017}) \bibinfo{pages}{6765--6816}.
\bibitem[{Falkner et~al.(2018)Falkner, Klein, and Hutter}]{falkner2018bohb}
\bibinfo{author}{S.~Falkner}, \bibinfo{author}{A.~Klein}, \bibinfo{author}{F.~Hutter},
\newblock \bibinfo{title}{Bohb: Robust and efficient hyperparameter optimization at scale},
\newblock in: \bibinfo{booktitle}{International Conference on Machine Learning}, \bibinfo{organization}{PMLR}, \bibinfo{year}{2018}, pp. \bibinfo{pages}{1437--1446}.
\bibitem[{Chen and Guestrin(2016)}]{chen2016xgboost}
\bibinfo{author}{T.~Chen}, \bibinfo{author}{C.~Guestrin},
\newblock \bibinfo{title}{Xgboost: A scalable tree boosting system},
\newblock in: \bibinfo{booktitle}{Proceedings of the 22nd acm sigkdd international conference on knowledge discovery and data mining}, \bibinfo{year}{2016}, pp. \bibinfo{pages}{785--794}.
\bibitem[{Ke et~al.(2017)Ke, Meng, Finley, Wang, Chen, Ma, Ye, and Liu}]{ke2017lightgbm}
\bibinfo{author}{G.~Ke}, \bibinfo{author}{Q.~Meng}, \bibinfo{author}{T.~Finley}, \bibinfo{author}{T.~Wang}, \bibinfo{author}{W.~Chen}, \bibinfo{author}{W.~Ma}, \bibinfo{author}{Q.~Ye}, \bibinfo{author}{T.-Y. Liu},
\newblock \bibinfo{title}{Lightgbm: A highly efficient gradient boosting decision tree},
\newblock \bibinfo{journal}{Advances in neural information processing systems} \bibinfo{volume}{30} (\bibinfo{year}{2017}).
\bibitem[{Lin et~al.(2017)Lin, Goyal, Girshick, He, and Doll{\'a}r}]{lin2017focal}
\bibinfo{author}{T.-Y. Lin}, \bibinfo{author}{P.~Goyal}, \bibinfo{author}{R.~Girshick}, \bibinfo{author}{K.~He}, \bibinfo{author}{P.~Doll{\'a}r},
\newblock \bibinfo{title}{Focal loss for dense object detection},
\newblock in: \bibinfo{booktitle}{Proceedings of the IEEE international conference on computer vision}, \bibinfo{year}{2017}, pp. \bibinfo{pages}{2980--2988}.
\bibitem[{Zhao et~al.(2019)Zhao, Nasrullah, and Li}]{zhao2019pyod}
\bibinfo{author}{Y.~Zhao}, \bibinfo{author}{Z.~Nasrullah}, \bibinfo{author}{Z.~Li},
\newblock \bibinfo{title}{Pyod: A python toolbox for scalable outlier detection},
\newblock \bibinfo{journal}{arXiv preprint arXiv:1901.01588}  (\bibinfo{year}{2019}).
\bibitem[{Shi et~al.(2022)Shi, Wong, Chai, Li, Chen, and Zeng}]{shi2022automaticclustering}
\bibinfo{author}{X.~Shi}, \bibinfo{author}{Y.~D. Wong}, \bibinfo{author}{C.~Chai}, \bibinfo{author}{M.~Z.-F. Li}, \bibinfo{author}{T.~Chen}, \bibinfo{author}{Z.~Zeng},
\newblock \bibinfo{title}{Automatic clustering for unsupervised risk diagnosis of vehicle driving for smart road},
\newblock \bibinfo{journal}{IEEE Transactions on Intelligent Transportation Systems} \bibinfo{volume}{23} (\bibinfo{year}{2022}) \bibinfo{pages}{17451--17465}.
\bibitem[{Shi et~al.(2020)Shi, Wong, Chai, and Li}]{shi2020automated}
\bibinfo{author}{X.~Shi}, \bibinfo{author}{Y.~D. Wong}, \bibinfo{author}{C.~Chai}, \bibinfo{author}{M.~Z.-F. Li},
\newblock \bibinfo{title}{An automated machine learning (automl) method of risk prediction for decision-making of autonomous vehicles},
\newblock \bibinfo{journal}{IEEE Transactions on Intelligent Transportation Systems} \bibinfo{volume}{22} (\bibinfo{year}{2020}) \bibinfo{pages}{7145--7154}.
\bibitem[{Arrieta et~al.(2020)Arrieta, D{\'\i}az-Rodr{\'\i}guez, Del~Ser, Bennetot, Tabik et~al.}]{arrieta2020explainable}
\bibinfo{author}{A.~B. Arrieta}, \bibinfo{author}{N.~D{\'\i}az-Rodr{\'\i}guez}, \bibinfo{author}{J.~Del~Ser}, \bibinfo{author}{A.~Bennetot}, \bibinfo{author}{S.~Tabik}, et~al.,
\newblock \bibinfo{title}{Explainable artificial intelligence (xai): Concepts, taxonomies, opportunities and challenges toward responsible ai},
\newblock \bibinfo{journal}{Information fusion} \bibinfo{volume}{58} (\bibinfo{year}{2020}) \bibinfo{pages}{82--115}.
\bibitem[{Frank~Hutter(2414)}]{ixautoml}
\bibinfo{author}{K.~E. Frank~Hutter, Marius~Lindauer}, \bibinfo{title}{ixautoml: Interactive and explainable automl}, \bibinfo{howpublished}{\url{https://www.automl.org/ixautoml/}}, \bibinfo{year}{Accessed 2024/1/4}.
\bibitem[{Academician~of CAS~Member(3719)}]{2021WSCE}
\bibinfo{author}{H.~W. Academician~of CAS~Member}, \bibinfo{title}{2021 world semiconductor conference:the chip challenges and opportunities in the post moore era}, \bibinfo{howpublished}{\url{http://www.wsc-expo.com/}}, \bibinfo{year}{Accessed 2023/7/19}.
\bibitem[{Todd~Edlund and of~semiconductor material~manufacturer Entegris(1115)}]{TODD}
\bibinfo{author}{E.~V.~P. Todd~Edlund}, \bibinfo{author}{C.~O.~O. of~semiconductor material~manufacturer Entegris}, \bibinfo{title}{A 1\% increase in yield can earn an additional 150 million dollars, highlighting the importance of semiconductor materials}, \bibinfo{year}{Accessed 2019/11/15}.
\bibitem[{in~2022 Gartner Magic~Quadrant(2022)}]{PAI}
\bibinfo{author}{A.~C. in~2022 Gartner Magic~Quadrant}, \bibinfo{title}{A platform that provides enterprise-level data modeling services based on machine learning algorithms to quickly meet your needs for data-driven operations.}, \bibinfo{year}{Accessed 2022}.
\bibitem[{Barnes(2015)}]{barnes2015azure}
\bibinfo{author}{J.~Barnes},
\newblock \bibinfo{title}{Azure machine learning},
\newblock \bibinfo{journal}{Microsoft Azure Essentials. 1st ed, Microsoft}  (\bibinfo{year}{2015}).
\bibitem[{Das et~al.(2020)Das, Ivkin, Bansal, Rouesnel, Gautier et~al.}]{das2020amazon}
\bibinfo{author}{P.~Das}, \bibinfo{author}{N.~Ivkin}, \bibinfo{author}{T.~Bansal}, \bibinfo{author}{L.~Rouesnel}, \bibinfo{author}{P.~Gautier}, et~al.,
\newblock \bibinfo{title}{Amazon sagemaker autopilot: a white box automl solution at scale},
\newblock in: \bibinfo{booktitle}{Proceedings of the fourth international workshop on data management for end-to-end machine learning}, \bibinfo{year}{2020}, pp. \bibinfo{pages}{1--7}.
\bibitem[{Furnari et~al.(2021)Furnari, Vattiato, Allegra, Milotta, Orofino, Rizzo, De~Palo, and Stanco}]{furnari2021ensembled}
\bibinfo{author}{G.~Furnari}, \bibinfo{author}{F.~Vattiato}, \bibinfo{author}{D.~Allegra}, \bibinfo{author}{F.~L.~M. Milotta}, \bibinfo{author}{A.~Orofino}, \bibinfo{author}{R.~Rizzo}, \bibinfo{author}{R.~A. De~Palo}, \bibinfo{author}{F.~Stanco},
\newblock \bibinfo{title}{An ensembled anomaly detector for wafer fault detection},
\newblock \bibinfo{journal}{Sensors} \bibinfo{volume}{21} (\bibinfo{year}{2021}) \bibinfo{pages}{5465}.
\bibitem[{Doke(2020)}]{doke2020datamining}
\bibinfo{author}{O.~Doke}, \bibinfo{title}{Data Mining for Enhancing Silicon Wafer Fabrication}, Ph.D. thesis, Dublin, National College of Ireland, \bibinfo{year}{2020}.
\bibitem[{Elfadel et~al.(2019)Elfadel, Boning, and Li}]{conceptdrift}
\bibinfo{author}{I.~M. Elfadel}, \bibinfo{author}{D.~S. Boning}, \bibinfo{author}{X.~Li}, \bibinfo{title}{Machine learning in VLSI computer-aided design}, \bibinfo{publisher}{Springer}, \bibinfo{year}{2019}.
\bibitem[{Watanabe et~al.(2023)Watanabe, Bansal, and Hutter}]{HyperparameterImportance}
\bibinfo{author}{S.~Watanabe}, \bibinfo{author}{A.~Bansal}, \bibinfo{author}{F.~Hutter},
\newblock \bibinfo{title}{Ped-anova: Efficiently quantifying hyperparameter importance in arbitrary subspaces},
\newblock \bibinfo{journal}{arXiv preprint arXiv:2304.10255}  (\bibinfo{year}{2023}).
\bibitem[{Z{\"o}ller et~al.(2023)Z{\"o}ller, Titov, Schlegel, and Huber}]{zoller2023xautoml}
\bibinfo{author}{M.-A. Z{\"o}ller}, \bibinfo{author}{W.~Titov}, \bibinfo{author}{T.~Schlegel}, \bibinfo{author}{M.~F. Huber},
\newblock \bibinfo{title}{Xautoml: A visual analytics tool for understanding and validating automated machine learning},
\newblock \bibinfo{journal}{ACM Transactions on Interactive Intelligent Systems} \bibinfo{volume}{13} (\bibinfo{year}{2023}) \bibinfo{pages}{1--39}.
\bibitem[{Sheikholeslami et~al.(2021)Sheikholeslami, Meister, Wang, Payberah, Vlassov, and Dowling}]{autoablation}
\bibinfo{author}{S.~Sheikholeslami}, \bibinfo{author}{M.~Meister}, \bibinfo{author}{T.~Wang}, \bibinfo{author}{A.~H. Payberah}, \bibinfo{author}{V.~Vlassov}, \bibinfo{author}{J.~Dowling},
\newblock \bibinfo{title}{Autoablation: Automated parallel ablation studies for deep learning},
\newblock in: \bibinfo{booktitle}{Proceedings of the 1st Workshop on Machine Learning and Systems}, \bibinfo{year}{2021}, pp. \bibinfo{pages}{55--61}.
\bibitem[{Moosbauer et~al.(2021)Moosbauer, Herbinger, Casalicchio, Lindauer, and Bischl}]{moosbauer2021explaining}
\bibinfo{author}{J.~Moosbauer}, \bibinfo{author}{J.~Herbinger}, \bibinfo{author}{G.~Casalicchio}, \bibinfo{author}{M.~Lindauer}, \bibinfo{author}{B.~Bischl},
\newblock \bibinfo{title}{Explaining hyperparameter optimization via partial dependence plots},
\newblock \bibinfo{journal}{Advances in Neural Information Processing Systems} \bibinfo{volume}{34} (\bibinfo{year}{2021}) \bibinfo{pages}{2280--2291}.
\bibitem[{Biedenkapp et~al.(2019)Biedenkapp, Marben, Lindauer, and Hutter}]{biedenkapp2019cave}
\bibinfo{author}{A.~Biedenkapp}, \bibinfo{author}{J.~Marben}, \bibinfo{author}{M.~Lindauer}, \bibinfo{author}{F.~Hutter},
\newblock \bibinfo{title}{Cave: Configuration assessment, visualization and evaluation},
\newblock in: \bibinfo{booktitle}{Learning and Intelligent Optimization: 12th International Conference, LION 12, Kalamata, Greece, June 10--15, 2018, Revised Selected Papers 12}, \bibinfo{organization}{Springer}, \bibinfo{year}{2019}, pp. \bibinfo{pages}{115--130}.
\bibitem[{Lee et~al.(2023)Lee, Kim, and Yu}]{lee2023semiconductor}
\bibinfo{author}{T.-E. Lee}, \bibinfo{author}{H.-J. Kim}, \bibinfo{author}{T.-S. Yu},
\newblock \bibinfo{title}{Semiconductor manufacturing automation},
\newblock in: \bibinfo{booktitle}{Springer Handbook of Automation}, \bibinfo{publisher}{Springer}, \bibinfo{year}{2023}, pp. \bibinfo{pages}{841--863}.
\bibitem[{Fisher et~al.(2019)Fisher, Rudin, and Dominici}]{fisher2019all}
\bibinfo{author}{A.~Fisher}, \bibinfo{author}{C.~Rudin}, \bibinfo{author}{F.~Dominici},
\newblock \bibinfo{title}{All models are wrong, but many are useful: Learning a variable's importance by studying an entire class of prediction models simultaneously.},
\newblock \bibinfo{journal}{J. Mach. Learn. Res.} \bibinfo{volume}{20} (\bibinfo{year}{2019}) \bibinfo{pages}{1--81}.
\bibitem[{Bergstra et~al.(2013)Bergstra, Yamins, Cox et~al.}]{bergstra2013hyperopt}
\bibinfo{author}{J.~Bergstra}, \bibinfo{author}{D.~Yamins}, \bibinfo{author}{D.~D. Cox}, et~al.,
\newblock \bibinfo{title}{Hyperopt: A python library for optimizing the hyperparameters of machine learning algorithms},
\newblock in: \bibinfo{booktitle}{Proceedings of the 12th Python in science conference}, volume~\bibinfo{volume}{13}, \bibinfo{organization}{Citeseer}, \bibinfo{year}{2013}, p.~\bibinfo{pages}{20}.
\bibitem[{Yoon et~al.(2018)Yoon, Jordon, and Schaar}]{yoon2018gain}
\bibinfo{author}{J.~Yoon}, \bibinfo{author}{J.~Jordon}, \bibinfo{author}{M.~Schaar},
\newblock \bibinfo{title}{Gain: Missing data imputation using generative adversarial nets},
\newblock in: \bibinfo{booktitle}{International conference on machine learning}, \bibinfo{organization}{PMLR}, \bibinfo{year}{2018}, pp. \bibinfo{pages}{5689--5698}.
\bibitem[{Zhai et~al.(2023)Zhai, Shi, and Zeng}]{zhai2023AutoML}
\bibinfo{author}{W.~Zhai}, \bibinfo{author}{X.~Shi}, \bibinfo{author}{Z.~Zeng},
\newblock \bibinfo{title}{Adaptive modelling for anomaly detection and defect diagnosis in semiconductor smart manufacturing: A domain-specific automl},
\newblock in: \bibinfo{booktitle}{2023 IEEE International Conference on Cybernetics and Intelligent Systems (CIS) and IEEE Conference on Robotics, Automation and Mechatronics (RAM)}, \bibinfo{organization}{IEEE}, \bibinfo{year}{2023}, pp. \bibinfo{pages}{198--203}.
\bibitem[{Shi et~al.(2022)Shi, Wong, Chai, Li, Chen, and Zeng}]{shi2022autocluster}
\bibinfo{author}{X.~Shi}, \bibinfo{author}{Y.~D. Wong}, \bibinfo{author}{C.~Chai}, \bibinfo{author}{M.~Z.-F. Li}, \bibinfo{author}{T.~Chen}, \bibinfo{author}{Z.~Zeng},
\newblock \bibinfo{title}{Automatic clustering for unsupervised risk diagnosis of vehicle driving for smart road},
\newblock \bibinfo{journal}{IEEE Transactions on Intelligent Transportation Systems} \bibinfo{volume}{23} (\bibinfo{year}{2022}) \bibinfo{pages}{17451--17465}.
\bibitem[{Nakazawa and Kulkarni(2018)}]{waferdefect}
\bibinfo{author}{T.~Nakazawa}, \bibinfo{author}{D.~V. Kulkarni},
\newblock \bibinfo{title}{Wafer map defect pattern classification and image retrieval using convolutional neural network},
\newblock \bibinfo{journal}{IEEE Transactions on Semiconductor Manufacturing} \bibinfo{volume}{31} (\bibinfo{year}{2018}) \bibinfo{pages}{309--314}.
\bibitem[{Chien et~al.(2007)Chien, Wang, and Cheng}]{chien2007data}
\bibinfo{author}{C.-F. Chien}, \bibinfo{author}{W.-C. Wang}, \bibinfo{author}{J.-C. Cheng},
\newblock \bibinfo{title}{Data mining for yield enhancement in semiconductor manufacturing and an empirical study},
\newblock \bibinfo{journal}{Expert Systems with Applications} \bibinfo{volume}{33} (\bibinfo{year}{2007}) \bibinfo{pages}{192--198}.
\bibitem[{Shi et~al.(2018)Shi, Wong, Li, and Chai}]{shi2018key}
\bibinfo{author}{X.~Shi}, \bibinfo{author}{Y.~D. Wong}, \bibinfo{author}{M.~Z.~F. Li}, \bibinfo{author}{C.~Chai},
\newblock \bibinfo{title}{Key risk indicators for accident assessment conditioned on pre-crash vehicle trajectory},
\newblock \bibinfo{journal}{Accident Analysis \& Prevention} \bibinfo{volume}{117} (\bibinfo{year}{2018}) \bibinfo{pages}{346--356}.
\bibitem[{Cheon et~al.(2019)Cheon, Lee, Kim, and Lee}]{8657760}
\bibinfo{author}{S.~Cheon}, \bibinfo{author}{H.~Lee}, \bibinfo{author}{C.~O. Kim}, \bibinfo{author}{S.~H. Lee},
\newblock \bibinfo{title}{Convolutional neural network for wafer surface defect classification and the detection of unknown defect class},
\newblock \bibinfo{journal}{IEEE Transactions on Semiconductor Manufacturing} \bibinfo{volume}{32} (\bibinfo{year}{2019}) \bibinfo{pages}{163--170}. \DOIprefix\doi{10.1109/TSM.2019.2902657}.
\bibitem[{Klar et~al.(2024)Klar, Ruediger, Schuermann, Gen, Glatt, Ravani, and Aurich}]{XRL202474}
\bibinfo{author}{M.~Klar}, \bibinfo{author}{P.~Ruediger}, \bibinfo{author}{M.~Schuermann}, \bibinfo{author}{G.~T. Gen}, \bibinfo{author}{M.~Glatt}, \bibinfo{author}{B.~Ravani}, \bibinfo{author}{J.~C. Aurich},
\newblock \bibinfo{title}{Explainable generative design in manufacturing for reinforcement learning based factory layout planning},
\newblock \bibinfo{journal}{Journal of Manufacturing Systems} \bibinfo{volume}{72} (\bibinfo{year}{2024}) \bibinfo{pages}{74--92}. \DOIprefix\doi{https://doi.org/10.1016/j.jmsy.2023.11.012}.
\bibitem[{Wang et~al.(2022)Wang, Li, Gao, and Zhang}]{XAI2022381}
\bibinfo{author}{J.~Wang}, \bibinfo{author}{Y.~Li}, \bibinfo{author}{R.~X. Gao}, \bibinfo{author}{F.~Zhang},
\newblock \bibinfo{title}{Hybrid physics-based and data-driven models for smart manufacturing: Modelling, simulation, and explainability},
\newblock \bibinfo{journal}{Journal of Manufacturing Systems} \bibinfo{volume}{63} (\bibinfo{year}{2022}) \bibinfo{pages}{381--391}. \DOIprefix\doi{https://doi.org/10.1016/j.jmsy.2022.04.004}.
\bibitem[{Bennett et~al.(1994)Bennett, Tobin, and Gleason}]{588270}
\bibinfo{author}{M.~Bennett}, \bibinfo{author}{K.~Tobin}, \bibinfo{author}{S.~Gleason},
\newblock \bibinfo{title}{Overview of automatic defect classification},
\newblock in: \bibinfo{booktitle}{Proceedings of 1994 IEEE/SEMI Advanced Semiconductor Manufacturing Conference and Workshop (ASMC)}, \bibinfo{year}{1994}, pp. \bibinfo{pages}{272--}. \DOIprefix\doi{10.1109/ASMC.1994.588270}.
\bibitem[{Lu et~al.(2020)Lu, Xu, and Wang}]{lu2020smart}
\bibinfo{author}{Y.~Lu}, \bibinfo{author}{X.~Xu}, \bibinfo{author}{L.~Wang},
\newblock \bibinfo{title}{Smart manufacturing process and system automation--a critical review of the standards and envisioned scenarios},
\newblock \bibinfo{journal}{Journal of Manufacturing Systems} \bibinfo{volume}{56} (\bibinfo{year}{2020}) \bibinfo{pages}{312--325}.
\bibitem[{Moghaddam et~al.(2018)Moghaddam, Cadavid, Kenley, and Deshmukh}]{moghaddam2018SM}
\bibinfo{author}{M.~Moghaddam}, \bibinfo{author}{M.~N. Cadavid}, \bibinfo{author}{C.~R. Kenley}, \bibinfo{author}{A.~V. Deshmukh},
\newblock \bibinfo{title}{Reference architectures for smart manufacturing: A critical review},
\newblock \bibinfo{journal}{Journal of manufacturing systems} \bibinfo{volume}{49} (\bibinfo{year}{2018}) \bibinfo{pages}{215--225}.
\bibitem[{Arinez et~al.(2020)Arinez, Chang, Gao, Xu, and Zhang}]{AIAgent}
\bibinfo{author}{J.~F. Arinez}, \bibinfo{author}{Q.~Chang}, \bibinfo{author}{R.~X. Gao}, \bibinfo{author}{C.~Xu}, \bibinfo{author}{J.~Zhang},
\newblock \bibinfo{title}{Artificial intelligence in advanced manufacturing: Current status and future outlook},
\newblock \bibinfo{journal}{Journal of Manufacturing Science and Engineering} \bibinfo{volume}{142} (\bibinfo{year}{2020}) \bibinfo{pages}{110804}.
\bibitem[{Tasias(2022)}]{FEextra2022integrated}
\bibinfo{author}{K.~A. Tasias},
\newblock \bibinfo{title}{Integrated quality, maintenance and production model for multivariate processes: a bayesian approach},
\newblock \bibinfo{journal}{Journal of Manufacturing Systems} \bibinfo{volume}{63} (\bibinfo{year}{2022}) \bibinfo{pages}{35--51}.
\bibitem[{Zhang et~al.(2023)Zhang, Wang, Zhou, Chang, Ma, Jing, Cheng, Ding, and Zhao}]{ZHANG2023102121}
\bibinfo{author}{C.~Zhang}, \bibinfo{author}{Z.~Wang}, \bibinfo{author}{G.~Zhou}, \bibinfo{author}{F.~Chang}, \bibinfo{author}{D.~Ma}, \bibinfo{author}{Y.~Jing}, \bibinfo{author}{W.~Cheng}, \bibinfo{author}{K.~Ding}, \bibinfo{author}{D.~Zhao},
\newblock \bibinfo{title}{Towards new-generation human-centric smart manufacturing in industry 5.0: A systematic review},
\newblock \bibinfo{journal}{Advanced Engineering Informatics} \bibinfo{volume}{57} (\bibinfo{year}{2023}) \bibinfo{pages}{102121}. \DOIprefix\doi{https://doi.org/10.1016/j.aei.2023.102121}.
\bibitem[{Love et~al.(2023)Love, Fang, Matthews, Porter, Luo, and Ding}]{AEI-XAI}
\bibinfo{author}{P.~E. Love}, \bibinfo{author}{W.~Fang}, \bibinfo{author}{J.~Matthews}, \bibinfo{author}{S.~Porter}, \bibinfo{author}{H.~Luo}, \bibinfo{author}{L.~Ding},
\newblock \bibinfo{title}{Explainable artificial intelligence (xai): Precepts, models, and opportunities for research in construction},
\newblock \bibinfo{journal}{Advanced Engineering Informatics} \bibinfo{volume}{57} (\bibinfo{year}{2023}) \bibinfo{pages}{102024}.
\bibitem[{Kruger et~al.(2011)Kruger, Shih, Hattingh, and Van~Niekerk}]{kruger2011agent}
\bibinfo{author}{G.~H. Kruger}, \bibinfo{author}{A.~J. Shih}, \bibinfo{author}{D.~G. Hattingh}, \bibinfo{author}{T.~I. Van~Niekerk},
\newblock \bibinfo{title}{Intelligent machine agent architecture for adaptive control optimization of manufacturing processes},
\newblock \bibinfo{journal}{Advanced Engineering Informatics} \bibinfo{volume}{25} (\bibinfo{year}{2011}) \bibinfo{pages}{783--796}.
\bibitem[{Shen et~al.(2006)Shen, Hao, Yoon, and Norrie}]{shen2006applications}
\bibinfo{author}{W.~Shen}, \bibinfo{author}{Q.~Hao}, \bibinfo{author}{H.~J. Yoon}, \bibinfo{author}{D.~H. Norrie},
\newblock \bibinfo{title}{Applications of agent-based systems in intelligent manufacturing: An updated review},
\newblock \bibinfo{journal}{Advanced engineering Informatics} \bibinfo{volume}{20} (\bibinfo{year}{2006}) \bibinfo{pages}{415--431}.
\bibitem[{McCann and Johnston(2008)}]{misc_secom_179}
\bibinfo{author}{M.~McCann}, \bibinfo{author}{A.~Johnston}, \bibinfo{title}{{SECOM}}, \bibinfo{howpublished}{UCI Machine Learning Repository}, \bibinfo{year}{2008}. \bibinfo{note}{{DOI}: https://doi.org/10.24432/C54305}.

\end{thebibliography}

\end{document}